\documentclass{ws-procs9x6}

\newcommand{\open}{{<\kern -0.3 em{\scriptscriptstyle )}}}
\newcommand{\de}{d}

\begin{document}

\title{Transversity and dihadron fragmentation
  functions\footnote{\uppercase{W}ork partially
supported by the \uppercase{A}lexander von \uppercase{H}umboldt \uppercase{F}oundation.}}

\author{Alessandro Bacchetta}

\address{Institut f{\"u}r Theoretische Physik, Universit{\"a}t Regensburg,\\
D-93040 Regensburg, Germany}

\author{Marco Radici}

\address{Dipartimento di Fisica Nucleare e Teorica, Universit\`{a} di Pavia, and\\
Istituto Nazionale di Fisica Nucleare, Sezione di Pavia, I-27100 Pavia, Italy}

\maketitle

\abstracts{
The observation of the quark transversity distribution requires another soft
object sensitive to the quark's transverse spin.
Dihadron fragmentation functions represent a convenient tool to analyze
partonic spin, which can influence the angular distribution of the two
hadrons. 
In particular, the so-called 
interference fragmentation functions can be used to probe transversity
both in semi-inclusive deep inelastic
scattering as well as proton-proton collisions. We discuss two
single-spin asymmetries sensitive to transversity in the these
two processes, at leading twist and leading order in $\alpha_S$.}

Transversity has received a lot of attention during this
conference.
Its 
measurement is on the agenda of several experimental
collaborations. However, it is still not clear what will turn out to be the
best way to access it. Different options will be explored in the
next few years: polarized Drell-Yan, polarized $\Lambda$
production, azimuthal asymmetries in pion production and, finally, azimuthal
asymmetries in two-pion production.\footnote{At the moment, data are available
  only for azimuthal asymmetries in pion production.\cite{Airapetian:2004tw}} 
Each one of these processes has some
advantages and drawbacks, making it necessary to explore all of them at the
same time. Here, we focus on  azimuthal
asymmetries in two-hadron production.

\begin{figure}[t]
\begin{tabular}{cc}
\epsfxsize=2.2in\epsfbox{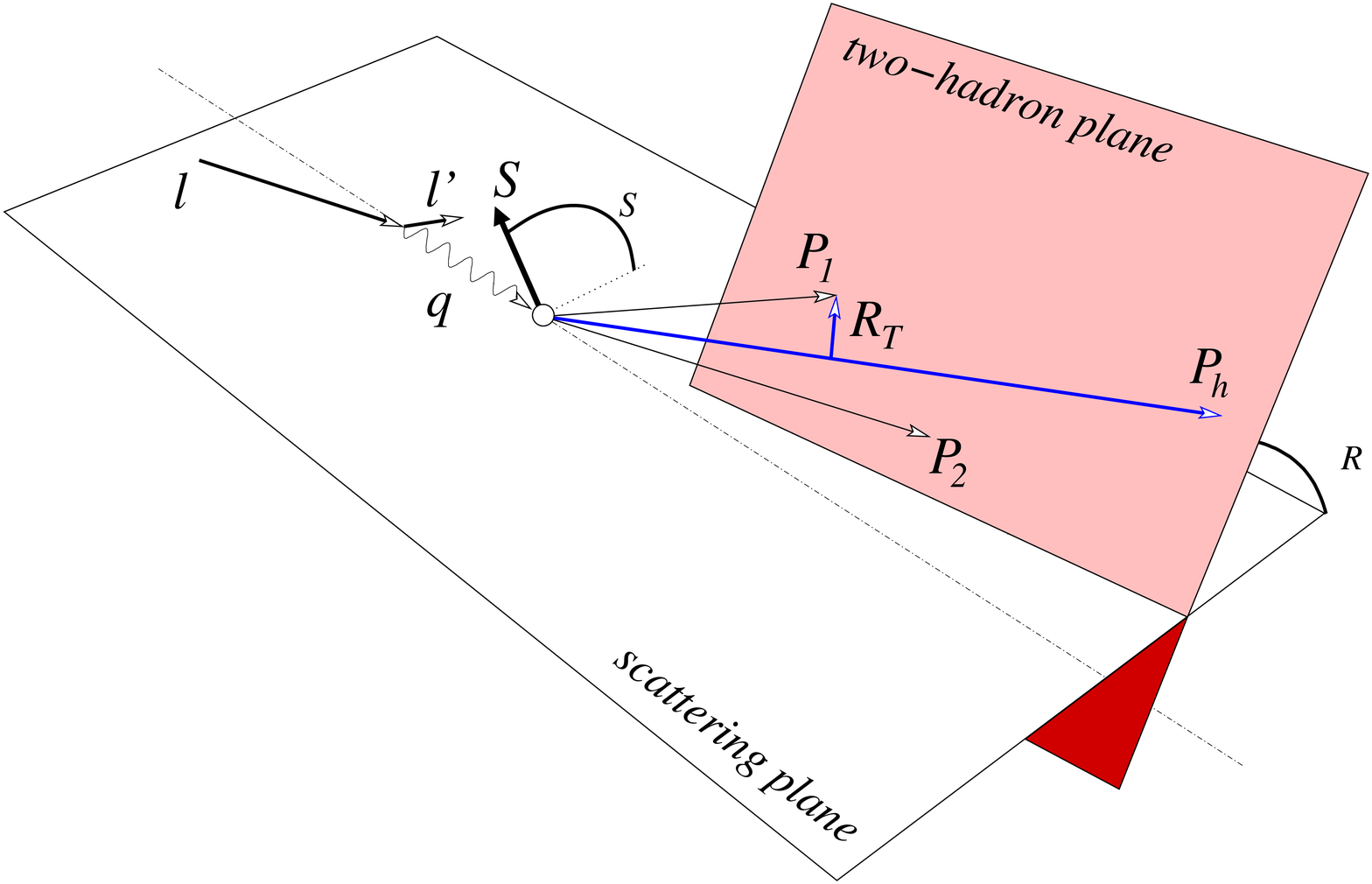}        
\put(-7,51){$\scriptstyle \varphi$}
\put(-91,81){$\scriptstyle \varphi$}
 &\epsfxsize=2.2in\epsfbox{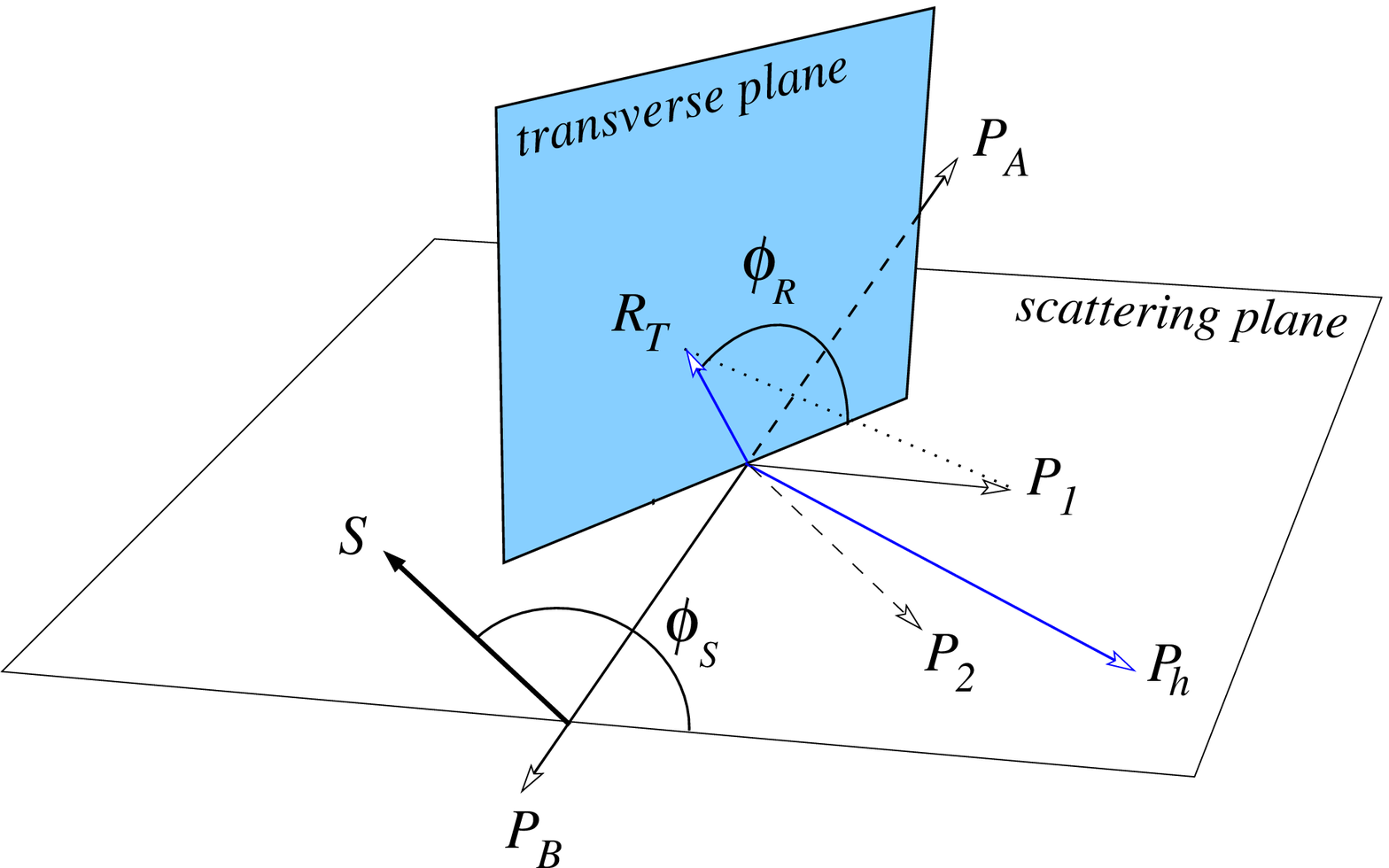}\\
(a) & (b)
\end{tabular}   
\caption{Angles involved in the measurement of the transverse 
single-spin asymmetry in (a) deep-inelastic production of two hadrons in the
current region, (b)
proton-proton collisions into two hadrons belonging to the same jet. 
\label{f:sidis}}
\end{figure}

First of all we consider leptoproduction of two hadrons in the current
fragmentation region, i.e.\ the process $lp \to l' (h_1 h_2) X$.
The outgoing hadrons have 
momenta $P_{1}$ and $P_{2}$, masses $M_1$ and $M_2$, and invariant mass $M_h$
(which must be much smaller than the virtuality of the photon, $Q$). 
We introduce the vectors $P_h=P_{1}+P_{2}$ and 
$R=(P_{1}-P_{2})/2$, i.e.\ the total and relative momenta of the pair, respectively. The angle 
$\theta$ is the angle between the direction of $P_1$
in the pair's center of mass and the direction of $P_h$ in the lab 
frame.\cite{Bacchetta:2002ux}
We introduce also the invariant
\begin{equation}
|{\bf R}|= \frac{1}{2} \, \sqrt{M_h^2 - 2(M_1^2+M_2^2) + (M_1^2-M_2^2)^2/M_h^2}.
\end{equation} 
Cross-sections are assumed to be differential in 
$\de M_h^2$, $\de\varphi_R^{}$, $\de z$, $\de x$, $\de y$, $\de\varphi_S^{}$, 
where $z$, $x$, $y$ are the usual scaling variables employed 
in semi-inclusive DIS, 
the azimuthal angles are defined so that (see Fig.~\ref{f:sidis}a)\footnote{The
  definition of the angles is consistent with the so-called Trento conventions.\cite{Bacchetta:2004jz}}
\begin{align} 
  \label{angle-def-1}
\cos \varphi_S &= 
  \frac{(\hat{\bf q}\times{\bf l})}{|\hat{\bf q}\times{\bf l}|}
  \cdot \frac{(\hat{\bf q}\times{\bf S})}{|\hat{\bf q}
     \times{\bf S}|}, 
& 
\sin \varphi_S &= 
  \frac{({\bf l} \times {\bf S}) \cdot \hat{\bf q}}{|\hat{\bf q}
     \times{\bf l}|\,|\hat{\bf q}\times{\bf S}|} , \\
\cos \varphi_R &=   
  \frac{(\hat{\bf q}\times{\bf l})}{|\hat{\bf q}\times{\bf l}|}
  \cdot \frac{(\hat{\bf q}\times{\bf R_T})}{|\hat{\bf q}
     \times{\bf R_T}|}, 
&
\sin \varphi_R 
&= 
  \frac{({\bf l} \times {\bf R_T}) \cdot \hat{\bf q}}{|\hat{\bf q}
     \times{\bf l}|\,|\hat{\bf q}\times{\bf R_T}|} , 
\end{align} 
where $\hat{\bf q} = {\bf q}/|{\bf q}|$ and ${\bf R_T}$ is the component of
$R$ perpendicular to $P_h$.

The best observable to be measured to attempt an extraction of transversity
with dihadron fragmentation functions is probably the transverse spin asymmetry~\cite{Radici:2001na}
\begin{equation}
A_{UT}(x,M_h^2,\varphi_R,\varphi_S)=
\frac{\int 
\de z\,\de y \, \de^6\!
  \sigma^{}_{UT}}
{\int 
\de z\,\de y \, \de^6\! \sigma^{}_{UU}}, 
\end{equation} 
where the unpolarized cross sections and the
transversely polarized cross section (difference) read
{\em up to leading twist only}\footnote{For subleading-twist corrections, see
  Refs.~\cite{Bacchetta:2003vn,Bacchetta:2004kn}. NLO $\alpha_S$ corrections
  have not been studied yet.}
\begin{align} 
\de^6\! \sigma^{}_{UU} &=\sum_a \frac{\alpha^2 e_a^2}{\pi Q^2 y}\,
     (1-y+y^2/2)\, f_1^a(x)  D_{1,oo}^a(z,M_h^2),  
\label{eq:crossOO} \\
\de^6\! \sigma^{}_{UT} &= -\sum_a \frac{\alpha^2 e_a^2}{4 Q^2 y} 
\frac{|{\bf R}|}{M_h} |{\bf S}_{\perp}^{}|
    (1-y) \sin(\varphi_R^{} + \varphi_S^{})
   h_1^a(x) H_{1,ot}^{\open a}(z,M_h^2). 
\end{align} 
Here, we assumed to integrate over the angle $\theta$. In the
unpolarized case, the function $D_{1,oo}$ represents the sum of all
resonances and backgrounds contributing to the two-pion spectrum, plus
possible interferences between pion pairs belonging to the same partial
waves. The function $H_{1,ot}^{\open}$ in the transversely polarized case represent an
interference between $s$ and $p$ waves. It should be therefore sizeable only
at values of the invariant mass where both waves contribute (e.g. in the
neighborhood of the $\rho$ resonance), although the precise invariant-mass
profile of this fragmentation function is at the moment
unknown.\cite{Jaffe:1998hf}
If information
about the distribution over $\theta$ is retained, transversity could be
coupled also to a pure $p$-wave fragmentation function, whose invariant-mass
profile can be expected to follow a Breit-Wigner distribution peaked at the
$\rho$ resonance.
Measurements of this kind of asymmetries are already in progress at
HERMES and COMPASS.\cite{paul}

Next we consider the production of two hadrons in the same jet in
proton-proton collisions, i.e.\ the process $p_A p_B \to (h_1 h_2) X$. 
The two protons have momenta $P_A$ and $P_B$. We assume
that only proton $B$ is polarized and has spin $S$.
The transverse momentum of the
hadron pair has to be large compared to all hadron masses and to the invariant
mass of the pair.
We use the same variables introduced before to describe the hadron pair.
Cross-sections are assumed to be differential in 
$\de \eta$, $\de |{\bf  P}_{h\perp}|$, $ \de \cos {\theta}$, 
$\de M_h^2$, $ \de \phi_{R}$, $\de \phi_{S}$, where
$\eta$ and ${\bf P}_{h\perp}$ are the pseudo-rapidity (defined with respect to
$P_A$) and the transverse
momentum of the hadron pair (i.e.\ of the sum of the
outgoing hadron's momenta) 
and the azimuthal angles are defined so that (see Fig.~\ref{f:sidis}b).
\begin{align} 
  \label{angle-def-2}
\cos \phi_{S} &= 
  \frac{(\hat{\bf P}_B \times {\bf P}_h)}{|\hat{\bf P}_B\times{\bf P}_h|}
  \cdot \frac{(\hat{\bf P}_B\times{\bf S})}{|\hat{\bf P}_B
     \times{\bf S}|},
&
\sin \phi_{S} &= 
  \frac{({\bf P}_h \times {\bf S}) \cdot \hat{\bf P}_B}{|\hat{\bf P}_B
     \times{\bf P}_h|\,|\hat{\bf P}_B\times{\bf S}|} , \\
\label{eq:azang3}
\cos \phi_{R} &= 
  \frac{(\hat{\bf P}_h \times {\bf P}_A)}{|\hat{\bf P}_h\times{\bf P}_A|}
  \cdot \frac{(\hat{\bf P}_h\times{\bf R})}{|\hat{\bf P}_h
     \times{\bf R}|},
&
\sin \phi_{R} &= 
  \frac{({\bf P}_A \times {\bf R}) \cdot \hat{\bf P}_h}{|\hat{\bf P}_h
     \times{\bf P}_A|\,|\hat{\bf P}_h\times{\bf R}|}, 
\end{align} 
where $\hat{\bf P} = {\bf P}/|{\bf P}|$. The asymmetry to be measured is~\cite{Bacchetta:2004it}
\begin{equation}
A_N (\eta, |{\bf P}_{h\perp}|, M_h^2, \phi_{R}, \phi_{S})
= 
\frac{
\de^5 \sigma_{UT}}{
\de^5 \sigma_{UU}} \; ,
\label{eq:AN}
\end{equation}
with
\begin{align} 
\de^5 \sigma_{UU}&=  2 \, |{\bf P}_{h\perp}| \! \sum_{\scriptscriptstyle
a,b,c,d}\int \frac{\de x_a
  \de x_b }{4 \pi^2 \bar{z}_c}  f_1^a(x_a) 
  f_1^b(x_b)  \frac{\de \hat{\sigma}_{ab \to cd}}{\de \hat{t}} 
D_{1,oo}^c(\bar{z}_c, M_h^2),
\label{eq:sigmaOO}
\\
\begin{split} 
\de^5 \sigma_{UT} &= 2 \, |{\bf P}_{h\perp}| \sum_{\scriptscriptstyle a,b,c,d}\,\frac{|{\bf R}|}{M_h}\, |{\bf S}_{T}|
\,
\sin{(\phi_{S}-\phi_{R})} 
\\ &  \quad \times 
\int \frac{\de x_a \de x_b }{16 \pi \bar{z}_c} \, f_1^a(x_a) \, 
h_1^b(x_b) \, \frac{\de \Delta \hat{\sigma}_{ab^\uparrow \to c^\uparrow
    d}}{\de \hat{t}} 
 \, H_{1,ot}^{\open c}(\bar{z}_c, M_h^2),
\end{split} 
\\
\bar{z}_c &= \frac{|{\bf{P}}_{h\perp}|}{\sqrt{s}} \frac{x_a e^{-\eta}
    + x_b e^{\eta}}{x_a x_b}
\end{align}  
and $\de \hat{\sigma}_{ab \to cd}$ denote the well-known unpolarized partonic
cross sections,
while $\Delta \hat{\sigma}_{ab^\uparrow \to c^\uparrow d}$ denote 
the cross sections (differences) with transversely polarized partons $b$ and
$c$, to be found in Ref.~\cite{Bacchetta:2004it}.
Measurement of this kind of asymmetry could be performed at RHIC.

Dihadron fragmentation functions could be measured by themselves also in 
$e^+e^-$ annihilation, for instance at BELLE~\cite{ralf} and BABAR, and thus
allow the exctraction of transversity from the above-mentioned asymmetries.

\section*{Acknowledgments}
The work of A.~B. and his participaton to the conference 
has been supported by the Alexander von Humboldt Foundation.


\begin{thebibliography}{ab}


\bibitem{Airapetian:2004tw}
A.~Airapetian {\it et al.}  [HERMES Collaboration],
arXiv:hep-ex/0408013; G.~Schnell, these proceedings.

\bibitem{Bacchetta:2002ux}
A.~Bacchetta and M.~Radici,
Phys.\ Rev.\ D {\bf 67} (2003) 094002

\bibitem{Bacchetta:2004jz}
A.~Bacchetta, U.~D'Alesio, M.~Diehl and C.~A.~Miller,
arXiv:hep-ph/0410050.

\bibitem{Radici:2001na}
M.~Radici, R.~Jakob and A.~Bianconi,
Phys.\ Rev.\ D {\bf 65} (2002) 074031


\bibitem{Bacchetta:2003vn}
A.~Bacchetta and M.~Radici,
Phys.\ Rev.\ D {\bf 69} (2004) 074026

\bibitem{Bacchetta:2004kn}
A.~Bacchetta and M.~Radici,
proceedings of DIS 2004,
arXiv:hep-ph/0407345.

\bibitem{Jaffe:1998hf}
R.~L.~Jaffe, X.~Jin and J.~Tang,
Phys.\ Rev.\ Lett.\  {\bf 80} (1998) 1166

\bibitem{paul}
P.~van der Nat, these proceedings; R.~Joosten, these proceedings.

\bibitem{Bacchetta:2004it}
A.~Bacchetta and M.~Radici,
Phys.\ Rev.\ D {\bf 70} (2004) 094032

\bibitem{ralf}
R.~Seidl, these proceedings.

\end{thebibliography}
\end{document}